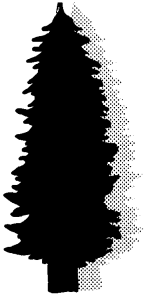

# Coming to Terms with Strongly Coupled Strings*

MICHAEL DINE
*Santa Cruz Institute for Particle Physics*
*University of California, Santa Cruz, CA 95064*

ABSTRACT

The holomorphy of the superpotential along with symmetries gives very strong constraints on any stringy non-perturbative effects. This observation suggests an approach to string phenomenology.

The enormous appeal of string theory is based entirely on the features which we understand in its weak coupling limit. Yet it has been clear for some time that if string theory describes nature it must be strongly coupled.[1] The problem is simply that any potential which might be generated (perturbatively or non-perturbatively) must vanish at weak coupling, i.e. at large values of the dilaton field, so any acceptable ground state must lie in a region where perturbation theory is not reliable. This is, at first sight, disappointing, since it is not clear that attractive features of the theory, such as its spectrum and symmetries should survive into the strong coupling regime. Moreover, one might despair of ever predicting anything from the theory.

Recently, there has been a marked improvement in our control over superstring dynamics. A remarkable amount of evidence has accumulated for a duality between strong and weak coupling in many instances, the so-called "S duality."[2,3] Yet, as wonderful as these connections are, they don't provide much help in dealing with the strong coupling problem I have just defined. For if a string model at very strong coupling is equivalent to some other string model at weak coupling, then the theory at very strong coupling suffers from the same instability as at weak coupling.

---

★ Work supported in part by the U.S. Department of Energy.
Paper presented at Strings 95: Future Perspectives in String Theory,
University of Southern California, Los Angles, CA, 13–18 March, 1995.

What really interests us, then, is some intermediate regime of coupling. A natural possibility to consider is that the coupling lies at some sort of self-dual or enhanced symmetry point of $S$ duality. However, this situation would be a very disappointing one, since the fine structure constants at this point would be of order one, unlike what we see in nature. The fact that the effective gauge couplings at high energies seem to be small would then be some kind of accident and any sort of unification a pure coincidence. Worse, we would not seem to have much hope of computing anything in the theory.

There is another possibility. It is known that the string perturbation expansion is not as convergent as field theory expansions.[4] Perhaps the string perturbation expansion is already not valid at values of the coupling which, to a field theorist, seem small.[5] At first sight, such a possibility is not obviously better than the prospect of very strong coupling. However, supersymmetry, coupled with certain symmetry properties of the theory, enormously constrain the structure of any possible stringy non-perturbative corrections under these circumstances. Indeed, all stringy corrections to the superpotential must be incredibly small. If this view is correct, the important stringy effects which (hopefully) stabilize the dilaton, explain the vanishing of the cosmological constant, and so on, must all be describable as $\mathcal{O}(1)$ modifications of the Kahler potential. The holomorphy of the superpotential and the gauge couplings also insure, under these circumstances, that many of the most attractive features of string theory survive into weak coupling:

1. The low energy theory is a supersymmetric theory with small, explicit soft breaking terms.
2. The spectrum of the theory is the same at strong coupling as at weak coupling.
3. Certain tree level relations among couplings receive only small corrections.

The keys to obtaining these results are the holomorphy of the superpotential and of the gauge coupling functions, and certain symmetries involving the dilaton. In Ref. 5, certain discrete gauged symmetries under which the dilaton transforms non-linearly played a crucial role. But it was also argued there that stringy non-perturbative effects are likely to respect a particular discrete axion shift, i.e. a shift of the dilaton supermultiplet,

$$S \to S + 2\pi i.$$

Here I am using a normalization of the dilaton multiplet different from that in most of the literature on duality:

$$S = \frac{8\pi^2}{g^2} + ia,$$

where $a$ is the axion field. With this normalization, an ordinary field theory instanton has an action proportional to $e^{-S}$. The assumption required to prove this symmetry is quite strong: all stringy non-perturbative effects, i.e. all effects which determine the low energy effective action near the string scale, are describable in terms of two dimensional field theories. But for those at this meeting, this symmetry is quite familiar – and eminently plausible: it is a subgroup of S duality.

Note that all of these statements apply to the Wilsonian effective action at some very high energy, i.e. an energy comparable to the string scale. In particular, infrared field theoretic effects will violate this argument. Gluino condensation in an $SU(N)$ group, for example, gives a contribution to the dilaton superpotential which behaves as $e^{-S/N}$. As explained in Ref. 5, this represents a spontaneous breaking of the symmetry. The gluino condensate has a phase which can take $N$ different values. A $2\pi$ shift in the axion is equivalent to a change of the gluino branch, and hence is still a symmetry.

If we accept the $2\pi$ periodicity for the axion, then, starting at weak coupling with a supersymmetric ground state, we can immediately make a number of important statements. First, we can bound the size of stringy non-perturbative effects (at weak coupling). Corrections to the superpotential necessarily behave as $e^{-nS}$. This is far smaller than many effects such as gluino condensation, which are visible in the low energy theory. Such stringy effects are still negligible when the coupling takes its "observed" value ($S \approx 200$). From this observation, it is also clear that the light spectrum cannot change. A change in the spectrum requires that some state with a mass of order the string scale come down to zero mass at some finite value of the coupling. But any such effect can be described in terms of the superpotential, and the superpotential is exponentially small.

This is not to say that there cannot be appreciable corrections to perturbative results. The Kahler potential is not constrained by holomorphy, and so, from this perspective, is basically arbitrary. The *potential* is given in terms of the Kahler potential and the superpotential, $W$, by

$$V = e^K \left[ \left( \frac{\partial W}{\partial \phi_i} + \frac{\partial K}{\partial \phi_i} W \right) g^{i\bar{j}} \left( \frac{\partial W^*}{\partial \phi_j^*} + \frac{\partial K}{\partial \phi_i^i} W \right) - 3|W|^2 \right] \quad (1)$$

(we have chosen units with the Planck scale set equal to one). Here the superpotential will have some form typical of gaugino condensation. With some "fiddling around," it is not difficult to construct Kahler potentials which yield a minimum of $V$ for large $S$ with vanishing cosmological constant. Of course, this vanishing of the cosmological constant requires incredible fine tuning; we have no solution to offer to this problem.

There are other important consequences of this viewpoint. First, corrections to Yukawa couplings (and other terms in the superpotential) will be very small. So relations which follow from the superpotential alone, but do not require any special properties of the Kahler potential, will hold even non-perturbatively. On the other hand, any predictions which depend on the detailed form of the lowest order Kahler potential are not likely to survive. As an example, some of the interesting ideas involving dilaton dominance which have been suggested to explain squark degeneracy cannot be employed in this framework.[6,7]

These simple observations suggest at least two directions for research. First, we can try to use holomorphy and symmetries as much as possible to understand the non-perturbative dynamics of string theories. Recently, Banks and I have demonstrated that certain string vacua have non-perturbative moduli in this way. We have also found cases where the system is repelled from or attracted to regions of higher symmetry in moduli space. Perhaps one will be able to make even stronger statements exploiting the recent understanding of duality between certain string theories. Second, one can try and do real phenomenology in this framework. One might examine particular string models with interesting features (e.g. three generations) and try to determine what sorts of statements can be made. Here one would see how far one could go by exploiting holomorphy and symmetries, within a framework of general "naturalness" considerations.[*] One would try to use only properties of the Kahler potential which follow from general symmetry considerations. This would amount to employing conventional notions of naturalness. For example, if some alignment of vev's is required to obtain a particular light spectrum, this should follow from symmetry considerations and not be simply assumed.

It is quite striking how far these sorts of simple symmetry arguments can take us. After all, we know very little about what string theory is non-perturbatively, much less about its dynamics. Yet perhaps there is, here, some inkling of how string theory might eventually make contact with nature.

## REFERENCES

1. M. Dine and N. Seiberg, *Phys. Lett.* **162B**, (1985) 299, and in *Unified String Theories*, M. Green and D. Gross, Eds. (World Scientific, 1986).
2. A. Font, L.E. Ibanez, D. Lust and F. Quevedo, *Phys. Lett.* **B249** (1990) 35; S.J. Rey, *Phys. Rev.* **D43** (1991) 526; A. Sen, *Nucl. Phys.* **B404** (1993) 109; Phys. Lett. **B303** (1993) 22; *Mod. Phys. Lett.* **A8** (1993) 2023; J. H. Schwarz and A. Sen, *Nucl. Phys.* **B411** (1994) 35.
3. See also the talks by C. Hull, E. Witten and A. Strominger at this meeting.
4. S. Shenker, RU-90-47 (1990).
5. T. Banks and M. Dine, *Phys. Rev.* **D50** (1994) 7454.
6. L.E. Ibanez and D. Lust, *Nucl. Phys.* **B382** (1992) 305.
7. V. Kaplunovsky and J. Louis, *Phys. Lett.* **B306** (1993) 269.

---

⋆ Of course one might object that the handling of the cosmological constant here is highly unnatural, so perhaps other "unnatural acts" should be permitted.